\documentclass[aps,groupedaddress,showpacs,epsfig,floats]{revtex4}
\usepackage[dvips]{graphicx}
\begin{document}

\title{Little-Parks effect in a superconducting loop with magnetic dot}
\author{ D. S. Golubovi\'{c}\footnote{Dusan.Golubovic@fys.kuleuven.ac.be}, W. V. Pogosov, M. Morelle and V. V. Moshchalkov}
\affiliation{Nanoscale Superconductivity and Magnetism Group,
Laboratory for Solid State Physics and Magnetism , K. U. Leuven,
Celestijnenlaan 200 D, B-3001 Leuven, Belgium}

\begin{abstract}
We have studied the nucleation of superconductivity in a
mesoscopic Al loop, enclosing magnetic dot with the perpendicular
magnetization. The superconducting phase boundary $T_{c}(B)$,
determined from transport measurements, is asymmetric with respect
to the polarity of an
 applied magnetic field. The maximum critical temperature has been
found for a finite applied magnetic field, which is antiparallel
to the magnetization of the dot. Theoretical phase boundary shows
a good agreement with the experimental data.
\end{abstract}
 \maketitle

A great deal of experimental and theoretical work has been devoted
to the investigation of mesoscopic superconductors (e.g. ref.
\cite {victor,vict1} and references therein). Over the past few
years {\it hybrid} superconductor/ferromagnet systems have
attracted a lot of attention
\cite{ja,pokro,rusi,misko,miskoa,miskob, danac,erdin,mjvb,martin}.
It has been revealed that a magnetic dot, embedded into a
superconductor, attracts a vortex, resulting in an enhanced
pinning when the magnetic moment $m$ of magnetic dot and an
applied magnetic field $B$ have parallel orientation
\cite{miskoa,mjvb,mjvbb}. Moreover, it was recently demonstrated
that perpendicularly magnetized magnetic dots could efficiently be
used not only to enhance the flux pinning, but also to tune the
nucleation of superconductivity \cite{martin}.

Theoretical studies of individual mesoscopic
superconductor/ferromagnet structures in the disk geometry were
carried out in the framework of the Ginzburg-Landau formalism,
with the emphasis on the vortex phases and transitions deep in the
superconducting state \cite{misko,miskob}. A mesoscopic
superconducting Al disk with a perpendicularly magnetized magnetic
dot was recently fabricated and its superconducting phase boundary
$T_{c}(B)$ was measured \cite{ja}. These measurements demonstrated
that the superconducting critical temperature is systematically
higher when the applied magnetic field is parallel to the
magnetization of the dot.
However, in this sample the superconducting phase boundary seems
to have been affected by the proximity effects between the
superconducting disk and magnetic dot.

In this paper the nucleation of superconductivity in a mesoscopic
superconducting Al loop, with a magnetic dot at the centre of the
opening, is investigated. Given the spatial separation of the
superconducting loop and magnetic dot, it is clear that the dot
interacts with the loop exclusively through the stray field and
proximity effects are absent. Therefore, any peculiar features in
the $T_{c}(B)$ phase boundary can undoubtedly be attributed to the
{\it magnetic} interaction between the superconducting condensate
and the stray field of the dot.


The sample was prepared by electron beam lithography in three
steps on a SiO$_{\rm 2}$ substrate using positive PMMA950K and the
corresponding co-PMMA electron beam resists. The lithography was
carried out in a JEOL 5600 scanning electron micrograph modified
for electron beam writing. The contacts pads, with the size of
$500\,{\rm \mu m}$, leads and alignment markers were patterned
first and $5\,{\rm nm}$ of Cr and $30\,{\rm nm}$ of Au were
thermally evaporated. Afterwards, the resist was removed with a
lift-off procedure, using warm acetone and ultrasonic agitation.
The sample was covered again
 with the e-beam resist. After the alignment, the
mesoscopic loop and contacts were patterned. The designed width of
the loop and contacts is $400\,{\rm nm}$, whereas the inner radius
of the loop is $600\,{\rm nm}$. In order to avoid proximity
effects, the mesoscopic contacts next to the loop are $22\,{\rm
\mu m}$ apart from the Cr/Au leads. $ 35\,{\rm nm}$ of Al was
thermally evaporated and a lift-off was carried out. In the final
step, the sample was yet again covered with the e-beam resists.
After a quite precise alignment procedure,  magnetic dot with the
radius of $200\,{\rm nm}$ was patterned. A magnetic dot consisting
of $2.5\,{\rm nm}$ Pd buffer layer and 10 bilayers of $0.4\,{\rm
nm}$ Co and $1\,{\rm nm}$ Pd was subsequently evaporated. The
resist was removed in a lift-off.
\begin{figure}[htb]
\centering
\includegraphics*[width=8cm]{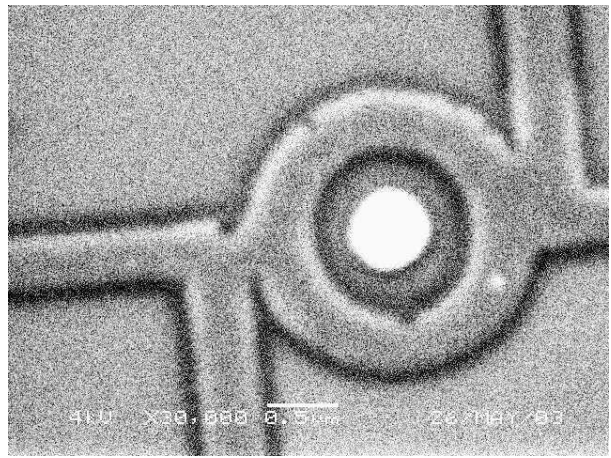}
\caption{A scanning electron micrograph of the structure.
\label{image}}
\end{figure}

Fig. \ref{image} shows an electron micrograph of the structure.
 The bright area at the centre is magnetic dot. The actual
 dimensions of the loop and magnetic dot were obtained
 from atomic force microscopy. The inner radius of the loop equals
 $R_{i}=0.55\,{\rm \mu m}$, outer radius is equal to $R_{o}=1.05\,{\rm \mu m}$, whereas
 the Co/Pd magnetic dot has the radius of $r_{d}=0.27\,{\rm \mu m}$. The mesoscopic contacts next
to the loop, used for measuring voltage, are closer to the loop
than initially designed. This effectively increases the outer
radius of the loop, since the supercurrent can spread into the
contacts, thereby affecting both the periodicity and background of
the phase boundary \cite{victor}.


The superconducting phase boundary $T_{c}(B)$ was obtained by
four-point transport measurements in a cryogenic setup at
temperatures down to $1.11\,$K, applying the magnetic field
perpendicularly to the sample surface. A transport current with
the rms of $150\,$nA and frequency $27.7\,$Hz was used. The
response was measured by a PAR124A lock-in amplifier. The phase
boundary was determined resistively by sweeping the magnetic field
at a very low rate, whilst keeping the temperature constant.
$T_{c}(B)$ curve was afterwards extracted from the
magnetoresistance using the resistive criterion $R_{n}/2$, where
$R_{n}$ is the resistance in the normal state. The temperature and
field steps were $500\,{\rm \mu K}$ and $5\,{\rm \mu T}$,
respectively, with the temperature stability $100\,{\rm \mu K}$.
In order to minimize the influence of possible temperature
gradients, the temperature was being monitored by two independent
temperature sensors placed at two different positions on the
sample holder.

The resistance of the sample at room temperature is $6.6\,{\rm
\Omega}$, the resistance in the normal state at low temperatures
is $R_{n}=2.5\,{\rm \Omega}$, whereas the maximum critical
temperature  $T_{cm}$ is $1.3422\,$K.

The multilayer Co/Pd magnetic dot provides a perpendicular
magnetization \cite{martin,gar}.
\begin{figure}[htb]
\centering
\includegraphics*[width=8cm]{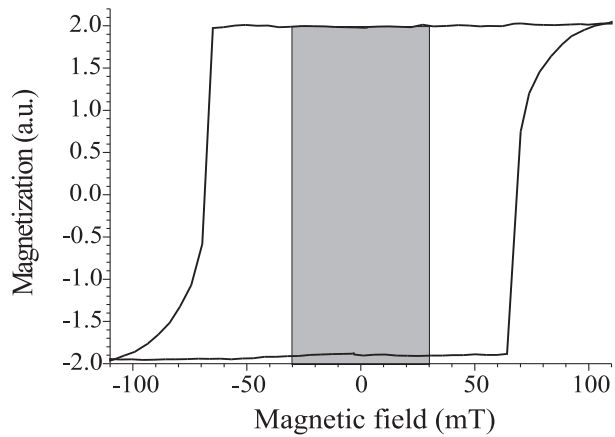}
\caption{The hysteresis loop of a co-evaporated Co/Pd plane film
obtained by the magneto-optical Kerr measurements at room
temperature. The shaded area indicates the range of the applied
magnetic field. \label{moke}}
\end{figure}
The hysteresis loop in the perpendicular magnetic field of the
co-evaporated plane film with the same composition, obtained by
the magneto-optical Kerr measurements at room temperature, is
presented in Fig. \ref{moke}. The shaded area indicates the range
in which the applied magnetic field was varied in the experiment.
The remanence is complete, whereas the coercive field is
approximately $70\, {\rm mT}$. When saturated, Co/Pd structures
patterned at the sub-micrometre scale are in the single-domain
state \cite{martin}.
\begin{figure}[htb]
\centering
\includegraphics*[width=8cm]{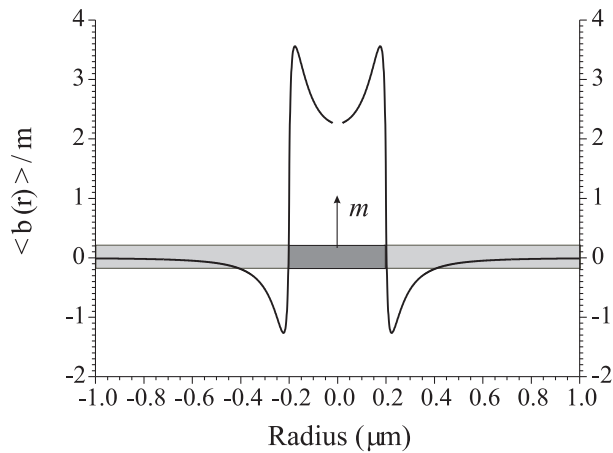}
\caption{The calculated stray field of the magnetic dot, averaged
over the thickness of the superconducting loop. The stray field is
normalized to the magnetization of the magnetic dot and was
calculated assuming a single domain structure. The darker shaded
area indicates the position of the dot, whereas the lighter shaded
area shows the position of the superconducting loop. The arrow
shows the direction of the magnetization. \label{stray}}
\end{figure}

Fig. \ref{stray} shows the calculated  stray field produced by a
Co/Pd magnetic dot, averaged over the thickness of the
superconducting loop. The stray field is normalized to the
magnetization of the magnetic dot. The darker shaded area shows
the position of the magnetic dot, whereas the lighter shaded areas
indicate the position of the superconducting loop. The arrow shows
the direction of the magnetization $m$ of the dot. The stray field
was calculated assuming a single-domain state of the magnetic dot.
Prior to the measurements the dot was magnetized perpendicularly
in $300\,$mT. As the applied magnetic field in our experiment
never exceeded $30\,$mT (Fig. \ref{moke}), it has been assumed
that the magnetization of the magnetic dot remains unaltered
during the measurements \cite{martin}.

The fluxes of the stray field (see Fig. \ref{stray}) through the
opening of the loop ($r<0.55\,{\rm \mu m}$) and loop itself ($0.55
< r < 1.05\,[{\rm \mu m}]$) have different orientations and, in
qualitative terms, there should be a competition between the
effects of these two fluxes on the superconducting condensate.
Assuming that the loop is sufficiently big compared to the
coherence length at zero temperature $\xi(0)$ ( $R_{i}, R_{o}>>\xi
(0)$, where $R_{i}$ and $R_{o}$ are the inner and outer radius of
the loop, respectively) the loop is similar to a stripe at
temperatures much lower than the maximum critical temperature the
system can attain $T_{cm}$ (we will be referring to these
temperatures as low temperatures). Clearly, the superconducting
state of the stripe depends only upon the flux value {\it within}
the stripe. In other words, at low temperatures the behaviour of
the loop is governed by the flux through the loop and not by the
flux in the opening of the loop. Therefore, at low temperatures
the shift of the $T_{c}(B)$ phase boundary along the $B$-axis is
expected to be in the direction of the dot magnetization. At
temperatures close to the maximum critical temperature (we will be
referring to these temperatures as high temperatures), the
direction of the shift of the $T_{c}(B)$ phase boundary is a
result of an interplay between the opposite fluxes. Note that the
superconducting phase boundary does not depend on the local
profile of the stray field in the opening of the loop, but on the
total flux through the opening generated by magnetic dot.

To describe quantitatively the $T_{c}(B)$ phase boundary the
Ginzburg-Landau theory has been used. The free energy of the
superconducting structure can be expressed as
\begin{eqnarray}
F= 2\pi \int_{R_{i}}^{R_{0}}\, rdr \left[ \alpha \psi + \beta \psi
^{3} + {\hbar ^{2}\over{4m}}\left(L-{2\pi\over{\Phi _{0}}}A
\right) \psi ^{2}+{\hbar
^{2}\over{4m}}\left({d\psi\over{dr}}\right)^{2} \right]
\label{eq1}
\end{eqnarray}
where $\psi$ is the modulus of the order parameter, $A$ is the
vector potential, an integer number $L$ stands the winding number
(vorticity) of the order parameter, $\Phi _{0}$ is the
superconducting flux quantum,  whereas $\alpha $ and $\beta$
represent the Ginzburg-Landau parameters. The integration is
performed over the area of the loop. In Eq. (\ref{eq1})
cylindrical symmetry of the order parameter near the phase
transition to the normal state has been taken into account. As the
thickness of sample is much smaller than $\xi(T)$ within the
temperature range of interest, there is no modulation of the order
parameter in the direction of the applied field. For this reason,
the stray field can be averaged out over the thickness of the
sample and the problem is reduced to the 1D case. Near the
superconducting phase boundary the total magnetic field is equal
to the sum of the stray field and an applied magnetic field.
Therefore, the vector potential may be expressed as
$A(r)=A_{s}(r)+Br/2$, where $A_{s}(r)$ is the vector potential of
the stray field and $B$ is an applied magnetic field.
We represent $\psi(r)$ as a Fourier expansion within the range
$R_{i}\leq r \leq R_{o}$ and consider the coefficients of the
expansion the variational parameters. The values of these
parameters can be calculated by a minimization of the free energy,
yielding the solution of the {\it full} Ginzburg-Landau system of
equations. The transition to the normal state corresponds to
$\psi=0$. We have found that for our sample and experimental
conditions, only the first term in the Fourier expansion is
important, that is, $\psi$ is constant within the sample, whereas
all other terms are negligibly small and practically do not affect
the position of the phase boundary.

\begin{figure}[htb]
\centering
\includegraphics*[width=8cm]{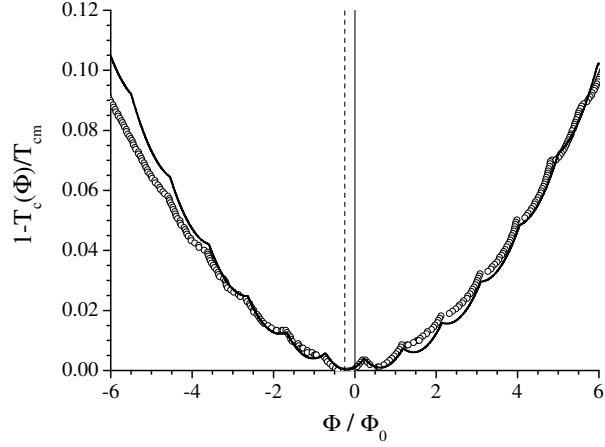}
\caption{The measured (open symbols) and calculated (solid line)
superconducting phase boundary given as $1-T_{c}(B)/T_{cm}$ versus
the normalized flux $\Phi / \Phi_{0}$ . Here $T_{cm}$ is the
maximum critical temperature and $\Phi_{0}$ is the superconducting
flux quantum. \label{tcb}}
\end{figure}

Fig. \ref{tcb} presents the superconducting phase boundary, given
as $1-T_{c}(B)/T_{cm}$ versus the normalized flux $\Phi
 / \Phi _{0}$.
$T_{cm}$ is the maximum critical temperature and $\Phi _{0}$ is
the superconducting flux quantum. Applied fields parallel to the
magnetization of the dot are taken as positive and vice versa. The
open symbols are experimental data, whereas the solid line
corresponds to the theoretical curve for $R_{i}= {0.55 \,\mu m}$,
and $r_{d}={0.27 \,\mu m}$. The coherence length at zero
temperature $\xi (T=0)$ has been used as a fitting parameter as it
depends upon the preparation procedure and can show moderate
variations even amongst samples prepared simultaneously. Due to
the influence of the mesoscopic contacts, the outer radius has to
be used as a fitting parameter, since the supercurrent is not
confined only by the ring, but can well flow into the mesoscopic
contacts, thus effectively increasing the outer radius. The best
fit has been obtained for the coherence length at zero temperature
$\xi (0)=96\,$nm and the outer radius $R_{o}=1130\,$nm. The latter
is quite close to the actual radius $R_{o}=1050\,$nm.

The theoretical $T_{c}(B)$ curve is in a good agreement with the
experimental data. The asymmetry, shift and periodicity of the
experimental data have been reproduced well. The asymmetry and the
shift can be explained by considering Eq. (\ref{eq1}). From this
equation it follows that the maximum critical temperature, for a
fixed $L$, is attained when the applied field is equal to
\begin{eqnarray}
B = -{4\Phi _{0}\over{\pi (R_{o}^{4}-R_{i}^{4}})}\int
_{R_{i}}^{R_{o}}\, r^{2}dr \left({2\pi
A\over{\Phi_{0}}}-{L\over{r}}\right) \label{eq2}
\end{eqnarray}
The corresponding critical temperature is

\begin{equation}
1-{T_{c}\over{T_{c\infty}}}={2\xi
(0)^{2}\over{R_{o}^{2}-R_{i}^{2}}}\left[\int _{R_{i}}^{R_{o}}\,
rdr \left({2\pi \over{\Phi_{0}}}A-{L\over{r}}
\right)^{2}-{4\over{R_{o}^{4}-R_{i}^{4}}}\left(\int
_{R_{i}}^{R_{o}} \,r^{2}dr
\left({2\pi\over{\Phi_{0}}}A-{L\over{r}} \right)\right)^{2}
 \right]
\label{eq3}
\end{equation}
where $T_{c\infty}$ stands for the bulk critical temperature of
Al. For the sake of simplicity, we assume that the stray field
does not change considerably over the width of the loop. Denoting
the average field in the opening of the loop and within the loop
as $B_{op}$ and $B_{l}$, respectively Eqs. (\ref{eq2}) and
(\ref{eq3}) reduce to

\begin{equation}
B=-B_{l}-{2\Phi_{0}\over{\pi(R_{0}^{2}+R_{i}^{2}})}\left({\pi
R_{i}^{2}(B_{l}-B_{op}) \over{\Phi_{0}}}-L\right) \label{eq4}
\end{equation}
and
\begin{equation}
1-{T_{c}\over{T_{c\infty}}}={2\xi(0)^{2}\over{R_{o}^{2}-R_{i}^{2}}}\left({\pi
R_{i}^{2}(B_{l}-B_{op})\over{\Phi_{0}}}-L\right)^{2}\left[{\rm
ln}({R_{0}\over{R_{i}}})-{R_{o}^{2}-R_{i}^{2}\over{R_{o}^{2}+R_{i}^{2}}}\right]
\label{eq5}
\end{equation}
The maximum critical temperature $T_{cm}$ can be found by the
minimization of the right-hand side of the Eq. (\ref{eq5}) with
respect to the integer $L$. Eq. (\ref{eq4}) gives the shift of the
$T_{c}(B)$ curve along the $B$-axis close to the $T_{cm}$, caused
by the presence of an additional magnetic field. This shift is a
sum of two contributions. The first term is just a compensation of
the stray field $B_{l}$ within the loop. The second term describes
the effect of the inhomogeneity of the stay field. This
contribution behaves rather nontrivially due to the discreteness
of $L$. For instance, when $0<\pi R_{i}^{2}(B_{op}-B_{l})<0.5\Phi
_{0}$, it follows from Eq. (5) that $L=0$, and the second
contribution to the shift is negative. If the size of the dot
and/or the intensity of the stray field are increased $0.5\Phi
_{0}<\pi R_{i}^{2}(B_{op}-B_{l})<\Phi _{0}$, $L$ switches  to $1$,
and the second contribution becomes positive. Physically, this
means that it is energetically more favorable for the system to
change the vorticity by $1$ in order to decrease the modulus of
the current in the loop. As a result, the current and the total
shift, defined by Eq. (4), can change the sign. Thus, we may
conclude, that the contribution to the shift, which is caused by
the stray field inhomogeneity, is a periodic function of
$B_{op}-B_{l}$ and vanishes in the case of a homogeneous field,
$B_{op}=B_{l}$. {\it The total shift of the $T_{c}(B)$ curve along
the $B$-axis close to the $T_{cm}$ can be of both signs for the
same orientation of magnetization of the dot.}

Our calculations have shown that at low temperatures the shift of
the phase boundary is controlled only by the average stray field
within the loop and is equal to -$B_{l}$, which is in accordance
with the preliminary qualitative analysis. If the additional
magnetic field were homogeneous, the shifts of the $T_{c}(B)$
phase boundary for low and high temperatures would have the same
direction. The inhomogeneity of the stay field gives rise to two
different and mutually independent shifts of the phase boundary in
the low and high temperature ranges, which, in turn, twist the
whole superconducting phase boundary, as presented in Fig.
\ref{tcb}.

We would like to note that the discrepancy between the theoretical
and the experimental data for low temperatures can be attributed
to the effect of contacts, since they change the geometry of the
sample, which cannot be entirely accounted for only by using the
outer radius of the loop as a fitting parameter.

To summarize, we have fabricated a mesoscopic superconducting loop
with a magnetic dot at the centre of the opening and investigated
the onset of superconductivity in this structure by measuring the
superconducting $T_{c}(B)$ phase boundary. The phase boundary has
been found to be asymmetric with respect to the polarity of an
applied magnetic field. The structure exhibits the maximum
critical temperature for a finite value of the applied magnetic
field, which is antiparallel to the magnetization of the magnetic
dot. The theoretical superconducting phase boundary, obtained in
the framework of the Ginzburg-Landau theory, is in a good
agreement with the experimental data. It has been demonstrated
that the inhomogeneous stray field shifts the $T_{c}(B)$ phase
boundary along the $B$-axis independently in the low and high
temperature ranges. In the limit of a thin loop with a large
radius, these shifts are determined by the average value of the
stray field in the opening of the loop and within the loop.

The authors would like to thank G. Rens for AFM measurements. This
work has been supported by the Belgian IUAP, the Flemish FWO  and
the Research Fund K. U. Leuven GOA/2004/02 programmes, as well as
by the ESF programme "VORTEX". W. V. P. acknowledges the support
from the Research Council of the K.U. Leuven and DWTC.


\begin{thebibliography}{20}
\bibitem{victor} V. V. Moshchalkov {\it et al.}, {\it Handbook of
Nanostructured Materials and Nanotechnology 3}, ed H. S. Nalwa
(Academic Press, San Diego, 2000).
\bibitem{vict1}V. V. Moshchalkov, V. Bruyndoncx and L. Van Look,
{\it Connectivity and Superconductivity}, ed J. Berger and J.
Rubinstein, (Springer-Verlag, Berlin 2000.)
\bibitem{ja}D. S. Golubovic, W. V. Pogosov, M. Morelle and V. V.
Moshchalkov, Appl. Phys. Lett., {\bf 83}, 1593 (2003).
\bibitem{pokro} V. L. Pokrovsky and H. Wei, cond-mat/0305153, 2003.
\bibitem{rusi} A Yu Aladyshkin {\it et al.}, cond-mat/0305551, 2003.
\bibitem{misko} M. V. Milo\v{s}evi\'{c}, S. V. Yampolskii and F.
M. Peeters, Phys. Rev. B, {\bf 66}, 024515 (2002).
\bibitem{miskoa}M. V. Milo\v{s}evi\'{c}, S. V. Yampolskii and F.
M. Peeters, Phys. Rev. B, {\bf 66}, 174519 (2002).
\bibitem{miskob}M. V. Milo\v{s}evi\'{c} and F. M. Peeters,
cond-mat/0211547, 2002.
\bibitem{danac}L. E. Helseth, Phys. Rev. B, {\bf 66}, 104508
(2002).
\bibitem{erdin} S. Erdin {\it et al.}, Phys. Rev. Lett., {\bf 88}, 1 (2002).
\bibitem{mjvb} M. J. Van Bael {\it et al.}, Phys. Rev. Lett., {\bf 86}, 1 (2001).
\bibitem{martin} M. Lange, M. J. Van Bael, Y. Bruynseraede and V. V. Moshchalkov,
Phys. Rev. Lett., {\bf 90}, 197006 (2003).
\bibitem{mjvbb} M . J. Van Bael {\it et al.}, Physica C, {\ 332}, 12
(2000).
\bibitem{gar}P. F. Carcia, J. Appl. Phys., {\bf 63}, 10 (1988).

\end{thebibliography}
\end{document}